\documentclass[10pt]{article}
\title{Non-time-orthogonality and Tests of Special Relativity}
\author{Robert D. Klauber\\1100 University Manor Dr., 38B, Fairfield, IA 52556, USA\\email: rklauber@netscape.net}
\date{December 29, 2001}
\usepackage {graphicx}
\usepackage {longtable}
\usepackage{hyperref}
\begin{document}
\maketitle


\bigskip

\begin{abstract}

An intriguing, and possibly significant, anomalous signal in the Brillet and 
Hall experiment is contrasted with a simple first order test of special 
relativity subsequently performed to discount that signal as spurious. 
Analysis of the non-time-orthogonal nature of the rotating earth frame leads 
to the conclusion that the latter test needed second order accuracy in order 
to detect the effect sought, and hence was not sufficient to discount the 
potential cause of the anomalous signal. The analysis also explains the 
results found in Sagnac type experiments wherein different media were placed 
in the path of the light beams.

\end{abstract}

\section{Introduction}

In 1981 Aspen\cite{Aspen:1981} showed that a persistent non-null test 
signal recorded by Brillet and Hall\cite{Brillet:1979} in the most modern 
Michelson-Morley type experiment performed to date would correspond to a 
test apparatus velocity of 363 m/sec. As the earth surface speed at the test 
location is approximately 355 m/sec, it was suggested that perhaps the 
rotating earth's surface velocity is somehow subtly different from the solar 
and galactic orbital velocities, which had long since been shown to yield 
null signals. To date there has been no other Michelson-Morley type test 
with sufficient accuracy to detect a possible effect from the earth surface 
velocity. 

In 1985, Byl et al\cite{John:1985} performed a first order test of special 
relativity which ostensibly showed no effect from the earth surface 
velocity, and led to the conclusion that the Brillet and Hall signal was 
erroneous. The Byl et al test was simple, clever, and based on an analysis 
that included certain, seemingly reasonable, kinematic and constitutive 
assumptions.

In this article we first review this test/analysis, then note that 
underlying assumptions are not in accord with that derived from 
non-time-orthogonal (NTO) analysis of rotating frames. It is then shown that 
if the NTO derivation result is correct, the effect of the earth surface 
velocity would be second, not first, order, and hence undetectable by the 
Byl et al experiment. This re-opens the question of the meaning and validity 
of the Brillet and Hall anomalous signal.

\section{``First Order'' Test of Special Relativity}
\label{sec:mylabel1}

\subsection{The Experiment}

Byl, Sanderse, and van der Kamp split a laser beam into two, directed one 
beam through air and one through water, and then combined the resultant 
beams to yield an interference pattern. According to their analysis 
(summarized below), the existence of an ether (privileged frame) would cause 
one of the beams to be slowed more than the other when both were aligned 
with the direction of absolute velocity. This would result in relative phase 
shifting and a movement of the fringes as the apparatus was turned.

\subsection{The Analysis}
\label{subsec:mylabel1}

Byl et al considered two cases: i) a direct Galilean transformation between 
the earth and ether frames with no ether drag, and ii) a Fresnel ether drag 
effect. We note that conclusions for each were similar and so only review 
the first case.

Assume a Galilean universe with light speed $c$ relative to the ether. The test 
apparatus moves towards the right with speed $v$. The top half of the original 
beam is directed toward the right through air while the bottom half is 
directed toward the right through water. Both travel a distance $D$ where the 
bottom beam emerges into air, and are at that point combined to form 
interference fringes. 

The travel time for the top beam is

\begin{equation}
\label{eq1}
t_{1} = {\frac{{D}}{{(c - v)}}}.
\end{equation}

The travel time for the bottom beam is

\begin{equation}
\label{eq2}
t_{2} = {\frac{{D}}{{u_{G}}} } = {\frac{{Dn}}{{(c - v)}}}
\end{equation}

\noindent
where $u_{G}$ is the presumed Galilean speed of light in water, and $n$ is the 
index of refraction of water.

The time difference is

\begin{equation}
\label{eq3}
T_{a} = t_{2} - t_{1} = {\frac{{D(n - 1)}}{{(c - v)}}}.
\end{equation}

When the apparatus is rotated 180$^{o}$ the difference in travel time 
becomes

\begin{equation}
\label{eq4}
T_{b} = {\frac{{D(n - 1)}}{{(c + v)}}}.
\end{equation}

The change in time difference is thus

\begin{equation}
\label{eq5}
\Delta T = T_{b} - T_{a} = {\frac{{2D(n - 1)vc}}{{c^{2} - v^{2}}}}.
\end{equation}

The displacement length of one wave to the other is then $c\Delta T$, leading 
to a displacement in $M$ wavelengths, or fringes, of

\begin{equation}
\label{eq6}
M = {\frac{{c\Delta T}}{{\lambda}} } = {\frac{{2D(n - 1)}}{{\lambda 
}}}{\frac{{vc}}{{c^{2} - v^{2}}}}
\end{equation}

\noindent
or

\begin{equation}
\label{eq7}
M = {\frac{{2D(n - 1)}}{{\lambda}} }{\frac{{v}}{{c}}} + O[(v / c)^{3}]
\end{equation}

Hence, the experiment was considered to have first order accuracy in $v/c$, be 
more sensitive than the (second order) Michelson-Morley experiment, and be 
capable of readily detecting any effect from the earth surface velocity. 

With the given apparatus, accuracy was deemed to be at least 20 times that 
needed to check the earth surface speed hypothesis. Repeated testing showed 
no detectable motion of the fringes.

\section{Index of Refraction and NTO Frames}
\label{sec:index}

\subsection{NTO Frames Theory}

Klauber\cite{Robert:1998}$^{,}$\cite{Robert:1999} has analyzed the NTO 
frame metric obtained when one makes a straightforward transformation from 
the lab to a rotating frame such as that of a relativistically spinning 
disk. Unlike other researchers\cite{For:1975} he has not assumed that it is 
then necessary to transform to locally time orthogonal frames. Instead he 
considered what phenomena would result if one proceeded using the NTO frame 
metric as a physically valid representation of the rotating frame.

Klauber found time dilation and mass-energy dependence on tangential speed 
$\omega r$ that is identical to the predictions of special relativity and the 
test data from numerous cyclotron experiments. He further found resolutions 
of several well known, and other not so well known, paradoxes inherent in 
the traditional analytical treatment of rotating frames. 

However, he also found some behavior that may seem somewhat unusual from a 
traditional relativistic standpoint, and which is difficult to verify 
experimentally. In particular, it was found that the 
physical\cite{All:1}$^{,}$\cite{Robert:1} (i.e., measured with standard 
instruments, not merely generalized coordinate value) velocity of light in 
vacuum in the circumferential direction on a rotating (NTO) frame is 
non-invariant, non-isotropic, and equals\cite{Ref:1}

\begin{equation}
\label{eq8}
u_{light,vacuum} \,\; = \;\,\,{\frac{{\pm c - r\omega}} {{\sqrt {1 - 
(r\omega )^{2} / c^{2}}}} } = {\frac{{\pm c - v}}{{\sqrt {1 - v^{2} / c^{2}} 
}}},
\end{equation}

\noindent
where $\omega $ is the angular velocity, $r$ is the radial distance from the 
center of rotation, $v=\omega r$, and the sign before $c$ depends on the 
propagation direction of the light ray.

This relationship leads to the prediction of a signal due to the earth 
surface speed $v$ precisely like that found by Brillet and Hall\cite{Ref:2}. 
For bodies in gravitational orbit (\ref{eq8}) does not hold, and $u_{light} = \pm 
c$. This is because such bodies are in free fall, and are essentially 
inertial, Lorentzian, time orthogonal (TO) frames. They are not subject to 
the idiosyncrasies of non-time-orthogonality, so their orbital speed would 
not result in a non-null Michelson-Morley signal.

The most relevant of Klauber's findings for the present purposes is the 
generalization of (\ref{eq8}) to an arbitrary velocity (other than for light)$.$ This, 
the transformation of physical tangential (circumferential) velocities 
between inertial and rotating frames, is\cite{Ref:3}

\begin{equation}
\label{eq9}
u\; = \;\,\,{\frac{{U - r\omega}} {{\sqrt {1 - (r\omega )^{2} / c^{2}}}} }
\end{equation}

\noindent
where lower case represents the rotating frame and upper case the inertial 
frame in which the rotation axis is fixed.

\subsection{Speed of Light in Water}
\label{subsec:speed}

As NTO frames have unique properties, it is not appropriate to simply assume 
that a beam of light in such a frame passing through a medium fixed in that 
frame would have speed ($c\pm v)/n$, as was presumed in (\ref{eq2}) and (\ref{eq4}). However, 
given what we know about NTO frames and Lorentz frames, we can determine 
what a rotating frame observer would see for a beam of light passing through 
a medium fixed in a Lorentz frame having the same instantaneous velocity as 
that observer. It may then seem reasonable to assume that this speed is the 
proper one to use in calculations such as those of section 
\ref{subsec:mylabel1}. Further support for this assumption may be 
found in the Appendix, which provides an alternative derivation for the 
results of this section.

Consider three frames: the rotating frame k, the non-rotating frame K in 
which the axis of rotation is fixed, and a Lorentz frame K$_{1}$, having 
velocity $v=\omega r$ (relative to K) in the direction of the instantaneous 
velocity of a point fixed in k. K$_{1}$ is sometimes called a ``tangent'' or 
a ``co-moving'' frame. Consider a medium such as water fixed in K$_{1}$ with 
a light beam passing through it in the direction of \textbf{\textit{v}}. As 
measured in K$_{1}$, a Lorentz frame, the speed of that beam is $c/n$ where $n$ is 
the index of refraction of the medium.

K and K$_{1}$ are both Lorentz frames and the transformation law for 
velocities between such frames tells us that the light beam in question as 
seen from K has speed

\begin{equation}
\label{eq10}
U = {\frac{{c / n + v}}{{1 + {\frac{{(c / n)v}}{{c^{2}}}}}}} = 
{\frac{{c}}{{n}}} + v - {\frac{{v}}{{n^{2}}}}\;\; + \;O[1 / c].
\end{equation}

Transformation of $U$ to the rotating frame is done using (\ref{eq9}) whereupon we find 
the circumferential speed of the light beam in a transparent medium is seen 
from the rotating frame to be

\begin{equation}
\label{eq11}
u = {\frac{{c}}{{n}}} - {\frac{{v}}{{n^{2}}}}\;\; + \;O[1 / c] = 
{\frac{{c}}{{n}}}\left( {1 - {\frac{{v}}{{cn}}} + O[1 / c^{2}]} \right).
\end{equation}

\section{Fringe Analysis Revisited}
\label{sec:fringe}

Repeating the analysis method of section \ref{subsec:mylabel1}, we 
find the travel time for the top beam (in air) from (\ref{eq8}) as

\begin{equation}
\label{eq12}
t_{1} = {\frac{{D\sqrt {1 - v^{2} / c^{2}}}} {{(c - v)}}} = {\frac{{D}}{{c - 
v}}} + O[1 / c^{3}] = {\frac{{D}}{{c}}}\left( {1 + {\frac{{v}}{{c}}}} 
\right) + O[1 / c^{3}].
\end{equation}

The travel time for the bottom beam with the light in water velocity of (\ref{eq11}) 
is

\begin{equation}
\label{eq13}
t_{2} = {\frac{{D}}{{u}}} = {\frac{{D}}{{(c / n)\left( {1 - v / cn + O[1 / 
c^{2}]} \right)}}} = {\frac{{Dn}}{{c}}}\left( {1 + {\frac{{v}}{{cn}}}} 
\right) + O[1 / c^{3}].
\end{equation}

The time difference where, importantly, terms in $v$/$c^{2}$ cancel is

\begin{equation}
\label{eq14}
T_{a} = t_{2} - t_{1} = {\frac{{D}}{{c}}}\left( {n - 1} \right) + O[1 / 
c^{3}].
\end{equation}

Ignoring higher order terms, one finds this result is independent of $v$, and 
so, through second order, it will also equal $T_{b}$, the time difference 
when the apparatus is turned 180\r{} ${\rm u}$ Therefore the fringe shift 
change

\begin{equation}
\label{eq15}
M = {\frac{{c\Delta T}}{{\lambda _{air}}} } = {\frac{{c(T_{b} - T_{a} 
)}}{{\lambda _{air}}} } = \quad O[1 / c^{2}]
\end{equation}

\noindent
is second, not first order. (Wavelength in air is used in (\ref{eq15}) since the 
standing wave interference pattern occurs in air after the bottom beam has 
emerged from the water.)

Thus, no matter which way the apparatus is turned, there will be no first 
order phase shift between water and air light beams. Hence, if k is the 
earth frame, there will be no observable change in fringe location in the 
experiment of section \ref{sec:mylabel1}.

\section{Sagnac Experiments with Transparent Media}
\label{sec:sagnac}

Post\cite{Post:1967} summarized results of several Sagnac type experiments, 
some of which analyzed the effect of fixing transparent media to the 
rotating apparatus in the paths of the oppositely directed light beams. In 
repeated testing, no interference fringe change was observed between the 
light beams in air and light beams in other media configurations. Change 
was, however, observed when the media was stationary while the apparatus 
rotated$^{} $\cite{Ref:4}.

The Galilean analysis leading to (\ref{eq7}) is in conflict with these results, as 
it would predict a first order fringe shift between tests with no media and 
those with media fixed to the rotating apparatus. A purely Lorentzian 
analysis, while predicting no such shift for different media fixed to the 
apparatus, would also predict no fringe change for variations solely in 
angular velocity. That such change does in indeed occur is an indisputable 
experimental fact. Unlike either the Galilean or the Lorentz frame analysis, 
the NTO frame analysis predicts fringe changes in accord with all of the 
cited experiments.

\subsection{Sagnac and Rotating Media}
\label{subsec:sagnac}

Tests with media fixed to the rotating apparatus are governed in NTO 
analysis by (\ref{eq15}), which represents the fringe shift difference between waves 
passing through air and through a medium along the same path in a rotating 
frame. To first order this is zero regardless of direction of velocity $v$. 
Hence, both the counterclockwise and clockwise beams have no first order 
retardation difference, resulting in no observable fringe shift prediction, 
and agreement with experiment.

\subsection{Sagnac and Non-rotating Media}

As noted, Sagnac tests in which a transparent medium is fixed in the lab 
exhibit fringe shifting between the no media and media cases. We analyze 
this as follows.

Analogously with (\ref{eq12}), the times of travel as measured on the rotating frame 
for the clockwise (cw) and counterclockwise (ccw) light beams in air around 
a circumference $D$ using NTO analysis [see (\ref{eq8})] where $v = \omega r$ are

\begin{equation}
\label{eq16}
t_{cw,air} = {\frac{{D}}{{u_{cw,air}}} } = {\frac{{D\sqrt {1 - v^{2} / 
c^{2}}}} {{(c - v)}}} = {\frac{{D}}{{c - v}}} + O[1 / c^{3}] \cong 
{\frac{{D}}{{c}}}\left( {1 + {\frac{{v}}{{c}}}} \right)
\end{equation}

\noindent
and

\begin{equation}
\label{eq17}
t_{ccw,air} = {\frac{{D}}{{u_{ccw,air}}} } = {\frac{{D\sqrt {1 - v^{2} / 
c^{2}}}} {{(c + v)}}} \cong {\frac{{D}}{{c}}}\left( {1 - {\frac{{v}}{{c}}}} 
\right).
\end{equation}

Times for cw and ccw light to travel the same path $D$ for glass stationary 
while the apparatus rotates are found by noting that light speed in the 
glass media in the lab is $c/n$. Then from (\ref{eq9}) with $U=c/n$ we have

\begin{equation}
\label{eq18}
t_{cw,glass} = {\frac{{D}}{{u_{cw,glass}}} } = {\frac{{D\sqrt {1 - v^{2} / 
c^{2}}}} {{c / n - v}}} = {\frac{{Dn}}{{c}}}\left( {{\frac{{1}}{{1 - nv / 
c}}}} \right) + O[1 / c^{3}] \cong {\frac{{Dn}}{{c}}}\left( {1 + 
{\frac{{nv}}{{c}}}} \right)
\end{equation}

\noindent
and

\begin{equation}
\label{eq19}
t_{ccw,glass} = {\frac{{D}}{{u_{ccw,glass}}} } = {\frac{{D\sqrt {1 - v^{2} / 
c^{2}}}} {{c / n + v}}} \cong {\frac{{Dn}}{{c}}}\left( {1 - 
{\frac{{nv}}{{c}}}} \right).
\end{equation}

The travel time differences between the glass and air beams for each of the 
cw and ccw directions are then

\begin{equation}
\label{eq20}
\Delta t_{cw} = t_{cw,glass} - t_{cw,air} \cong {\frac{{D}}{{c}}}\left( {(n 
- 1) + (n^{2} - 1){\frac{{v}}{{c}}}} \right)
\end{equation}

\noindent
and

\begin{equation}
\label{eq21}
\Delta t_{ccw} = t_{ccw,glass} - t_{ccw,air} \cong {\frac{{D}}{{c}}}\left( 
{(n - 1) - (n^{2} - 1){\frac{{v}}{{c}}}} \right).
\end{equation}

The difference in reduction of arrival times\cite{One:1} between the cw and 
ccw beams is

\begin{equation}
\label{eq22}
\Delta t_{cw / ccw} = \Delta t_{cw} - \Delta t_{ccw} \cong 2D(n^{2} - 
1){\frac{{v}}{{c^{2}}}},
\end{equation}

\noindent
so the fringe shift location change between the air and stationary glass 
tests is

\begin{equation}
\label{eq23}
M = {\frac{{c\Delta t_{cw / ccw}}} {{\lambda _{air}}} } \cong 
{\frac{{2D}}{{\lambda _{air}}} }(n^{2} - 1){\frac{{v}}{{c}}}.
\end{equation}

Hence NTO analysis of stationary media placed in the Sagnac experiment light 
paths predicts a first order fringe shift increase in agreement with the 
test of Dufour and Prunier\cite{Dufour:1942} noted by Post\cite{Ref:5}.

\section{Summary and Conclusions}
\label{sec:summary}

The persistent non-null signal found in the Brillet and Hall experiment, the 
most modern and accurate Michelson-Morley type experiment to date, appears 
to correlate with the earth surface velocity. Non-time-orthogonal analysis 
of the rotating frame of the earth predicts this signal and implies that it 
can not be deemed spurious on the basis of the subsequent first order test 
carried out by Byl et al.

Sagnac type experiments carried out with various transparent media in the 
path of the light beams yield results that are in harmony with the 
predictions of NTO frame analysis, but are at variance with Lorentzian and 
Galilean frame analyses.

NTO analysis is ultimately consonant with the theory of relativity. Due to 
the idiosyncratic nature of NTO frames, however, it makes some predictions 
that do not seem traditionally relativistic.\cite{Ref:6}

\appendix

\section*{Appendix: Light Speed in Rotating Medium}

This appendix provides an alternative derivation to (\ref{eq11}). Some knowledge of 
NTO frames (see reference \cite{Robert:1998}) is assumed.

Figure 1 depicts the radial direction and time axes at a point fixed in the 
rotating frame k for both the rotating frame and a Lorentz frame K$_{1}$ 
having velocity equal to the tangential velocity of that point. The 
$X_{1}$ axis of K$_{1}$ is aligned with the radial axis of k. Event A has 
coordinates (\textit{ct} =0, $r=r_{o}$, $\phi $=0, $z$=0) in k and (\textit{cT}$_{1}$=0, $X_{1}$=0, 
$Y_{1}$=0, $Z_{1}$=0) in K$_{1}$. All distances and times are small. Both the 
$r$ axis in k and the $X_{1}$ axis in K$_{1}$ are orthogonal to their respective 
time axes. Only first order effects are considered, so any time dilation or 
Lorentz contraction effects are ignored. Hence, the coordinate time $t$ in k 
is, for present purposes, effectively equal to the time on local standard 
clocks in k, as well as time $T_{1}$ in K$_{1}$.

Each frame carries a tube of water aligned in the radial direction as well 
as standard clocks and measuring rods to determine both the speed of a light 
beam in water and that of a parallel light beam in air. (``Air'' is 
considered equivalent to ``vacuum'' here.) The light beam in air travels 
spacetime path AP from event A to event P. The light beam in water travels 
from event W to event P where it interferes constructively with the air 
light beam AP.

As all velocities are perpendicular to the spatial axes shown in Figure 1 of 
both k and K$_{1}$, we can assume paths AP and WP are identical in both 
frames. More specifically, light in the rotating tube of water travels at 
the same speed as light in the Lorentz frame tube of water, i.e., WP is the 
path of light through water in both frames.

The speed of light in air for both frames is $c$. The speed of light in water 
for both frames must be $c/n$, where $n$ is the index of refraction in water for a 
Lorentz frame. The wave amplitude peak for the water beam that arrives at P 
must be emitted (event W) before the wave amplitude peak of the air beam 
(emitted at A) so that they constructively interfere at P.

Now consider turning the apparatus in both frames until the tubes are 
pointed in the circumferential direction, $r\phi $ axis in k and $Y_{1}$ in K, 
as shown in Figure 2. The Lorentz frame remains unchanged in all regards. 
Time is still orthogonal to the $Y_{1}$ axis and all phenomena (time and 
location of emissions and subsequent constructive interference events) are 
identical. The rotating frame along the circumferential axis is, however, 
not time orthogonal (as described in reference \cite{Robert:1998}.) The 
degree of non-time-orthogonality in the tube direction increases as the 
tubes are turned, reaching a maximum in the circumferential direction.

\begin{figure}[htbp]
\includegraphics[bbllx=0.26in,bblly=0.13in,bburx=5.67in,bbury=3.10in,scale=1.00]{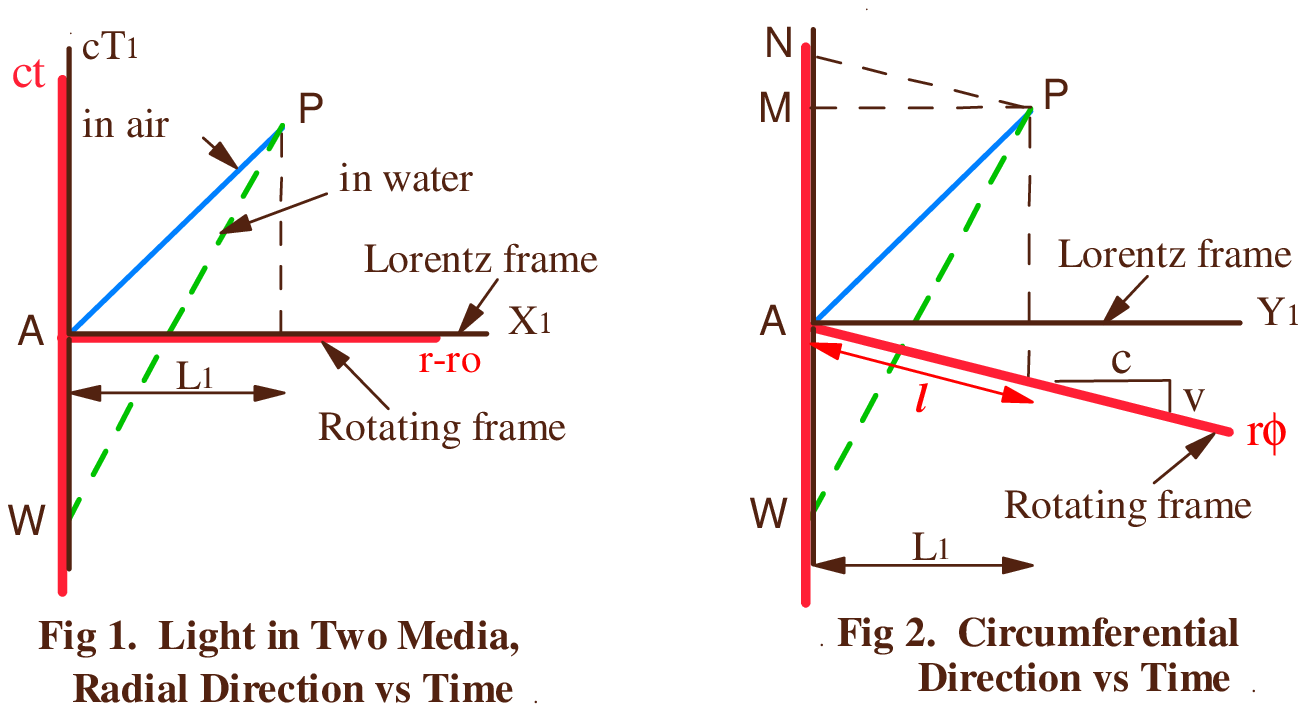}
\label{fig1}
\end{figure}

\bigskip

The NTO frame treatment of the aforementioned reference analyzes identical 
events, but due to the NTO nature of the frame, measured quantities such as 
time, distance, and light speed between those events differ from that of TO 
frames. We assume a similar stance with regard to the light beam passing 
through water. That is, there appears to be no constitutive reason why the 
path WP of a light beam in water in the Lorentz frame K$_{1}$ should differ 
from that of a light beam in water in the rotating frame k. The emitting 
event W and the constructive interference event P should be the same.

The issue then is to determine the speed of the light beam in water as 
measured in k. The first step in doing this is to note that the time of 
event P differs in the two frames. From Figure 2 it is seen to equal to the 
value on the time axis of event M in K$_{1}$, and to that of event N in k. 
We know the speed of light in water in the Lorentz frame K$_{1}$ is

\begin{equation}
\label{eq24}
V_{water,K1} = {\frac{{c}}{{n}}} = {\frac{{L_{1}}} {{\Delta t_{WM}}} }.
\end{equation}

The speed of the same light beam as seen in the rotating frame k is

\begin{equation}
\label{eq25}
v_{water,k} = {\frac{{l}}{{\Delta t_{WN}}} } \cong {\frac{{L_{1}}} {{\Delta 
t_{WM} + \Delta t_{MN}}} }.
\end{equation}

\noindent
where the RHS results because to first order $l \cong L_{1} $. From the slope 
--\textit{v/c= -$\omega $}\textit{${\rm g}$}$/c$ of NP, we find

\begin{equation}
\label{eq26}
c\Delta t_{MN} = L_{1} {\frac{{v}}{{c}}}.
\end{equation}

Using this in (\ref{eq25}) along with (\ref{eq24}), we find

\begin{equation}
\label{eq27}
v_{water,k} \cong {\frac{{L_{1}}} {{\Delta t_{WM}}} }\left( {1 - 
{\frac{{L_{1}}} {{\Delta t_{WM}}} }{\frac{{v}}{{c^{2}}}}} \right) = 
{\frac{{c}}{{n}}}\left( {1 - {\frac{{v}}{{cn}}}} \right),
\end{equation}

\noindent
the same relationship (where approximately equal means to zeroth order in 
1/$c)$ as (\ref{eq11}).


\begin{thebibliography}{13}
\bibitem{Aspen:1981} H. Aspen, ``Laser interferometry experiments on light speed anisotropy,'' \textit{Phys. Lett.,} \textbf{85A}(8,9), 411-414 (1981).
\bibitem{Brillet:1979} A. Brillet and J. L. Hall, ``Improved laser test of the isotropy of space,'' \textit{Phys. Rev. Lett}., \textbf{42}(\ref{eq9}), 549-552 (1979).
\bibitem{John:1985} John Byl, Martin Sanderse, and Walter van der Kamp, ``Simple first-order test of special relativity,'' \textit{Am. J. Phys.}, \textbf{53}(\ref{eq1}), 43-45 (1985).
\bibitem{Robert:1998} Robert D. Klauber, ``New perspectives on the relatively rotating disk and non-time-orthogonal reference frames'', \textit{Found. Phys. Lett.} \textbf{11}(\ref{eq5}), 405-443 (1998).
\bibitem{Robert:1999} Robert D. Klauber, ``Comments regarding recent articles on relativistically rotating frames'', \textit{Am. J. Phys.} \textbf{67}(\ref{eq2}), 158-159, (1999).
\bibitem{For:1975} For example, Adler, R., Bazin, M., and Schiffer, M., \textit{Introduction to General Relativity} (McGraw-Hill, New York, 1975), 2$^{nd}$ ed., pp. 121-122.
\bibitem{All:1} All quantities (such as the speed of light) used in all equations in this article are physical values, i.e., they equal what an experimentalist would measure with standard instruments. They are not coordinate values, which vary with the coordinate system used, but physical components, which are unique within a given frame and correspond to actual measured quantities. See ref. \cite{Robert:1}.
\bibitem{Robert:1} Robert D. Klauber, ``Physical components, coordinate components, and the speed of light'', gr-qc/0105071.
\bibitem{Ref:1} Ref. \cite{Robert:1998}, Section 4.2.4, pg. 425, eq. (\ref{eq19}) modified by the time dilation factor discussed in the subsequent paragraphs therein to yield physical velocity., and pg. 430, eq. (33).
\bibitem{Ref:2} Ref. \cite{Robert:1998}, Section 6. pp. 434-436.
\bibitem{Ref:3} Ref. \cite{Robert:1998}, Section 4.2.4, pp. 424-425. See eq. (\ref{eq18}) modified by the factor in the subsequent paragraph therein to yield physical velocity.
\bibitem{Post:1967} E. J. Post, ``Sagnac Effect'', \textit{Rev. Mod. Phys.}, \textbf{39}(\ref{eq2}), pp. 475-493 (1967).
\bibitem{Ref:4} Ref. \cite{Post:1967}, pp. 477-478 summarizes the effects of media on Sagnac test fringing. Note that fringe shift change is defined as the change in fringe location with respect to the stationary interferometer fringe location [pg. 476 before eq. (\ref{eq1})]. It is \textit{not} affected by the change in wavelength due to the insertion of transparent media in the light path. This is because the light waves emerge from the transparent media prior to forming a standing wave interference pattern, i.e., the interference fringing occurs in air. Any retardation that effects both waves to the same degree will have no effect on the standing wave interference pattern. That is, both traveling waves (that form the standing wave) arriving later by the same $\Delta $\textit{t} is equivalent to looking at the two original waves $\Delta $\textit{t} later in time. But the fringe pattern does not change with time. For example, in the non-rotating apparatus, insertion of a transparent medium in the paths of the light waves (such that both waves pass through the same length of the medium) will have no effect on the fringing. A change in location of interference fringes indicates one wave was retarded by a different amount than the other.
\bibitem{One:1} One notes that both light beams are slowed the same amount by the zeroth order (independent of \textit{v =} $\omega $\textit{r}) term in both (\ref{eq20}) and (\ref{eq21}). This term will have no effect on the location of the fringes. (See endnote \cite{Ref:4}.)
\bibitem{Dufour:1942} Dufour, A. and Prunier, F., \textit{J. Phys. Radium}, 8$^{th}$ Ser., \textbf{3}, 153 (1942).
\bibitem{Ref:5} Ref. \cite{Post:1967}, pg. 478.
\bibitem{Ref:6} Ref. \cite{Robert:1998}, Section 4.2, pp. 423-425.
\end{thebibliography}
\end{document}